\documentclass[aps,prl,english,10pt,superscriptaddress,twocolumn,float,floatfix]{revtex4-2}  
\usepackage{amsmath}
\usepackage{amsthm}
\usepackage{amssymb}
\usepackage{amsfonts}
\usepackage{bbm}
\usepackage{epsfig}
\usepackage{times}
\usepackage{cancel}
\usepackage{babel}
\usepackage{color}
\usepackage{hyperref}
\usepackage{framed}
\usepackage{changes}
\usepackage{graphicx}
\usepackage{braket}
\usepackage{orcidlink}

\begin{document}
	\title{Scaling at the OTOC Wavefront:   Free versus chaotic models}
	\author{Jonathon Riddell}
	\affiliation{Department of Physics \& Astronomy, McMaster University
		1280 Main St.  W., Hamilton ON L8S 4M1, Canada.}
		\author{Wyatt Kirkby}
	\affiliation{Department of Physics \& Astronomy, McMaster University
		1280 Main St.  W., Hamilton ON L8S 4M1, Canada.}
		\author{D. H. J. O'Dell\,\orcidlink{0000-0003-1463-1527}}
	\affiliation{Department of Physics \& Astronomy, McMaster University
		1280 Main St.  W., Hamilton ON L8S 4M1, Canada.}
        \author{Erik S. S{\o}rensen\,\orcidlink{0000-0002-5956-1190}}
	\affiliation{Department of Physics \& Astronomy, McMaster University
		1280 Main St.  W., Hamilton ON L8S 4M1, Canada.}
	
	\date{\today}

\begin{abstract}
Out of time ordered correlators (OTOCs) are useful tools for investigating foundational questions such as thermalization in closed quantum systems because they can potentially distinguish between integrable and nonintegrable dynamics.  Here we discuss the properties of wavefronts of OTOCs by focusing on the region around the main wavefront at $x=v_{B}t$, where $v_{B}$ is the butterfly velocity.  Using a Heisenberg spin model as an example, we find that the leading edge of a propagating Gaussian with the argument $-m(x)\left(  x-v_B t \right)^2
+b(x)t$ gives an excellent fit to the region around $x=v_{B}t$ for both the free and chaotic cases. However, the scaling in these two regimes is very different: in the free case the coefficients $m(x)$ and $b(x)$ have an inverse power law dependence on $x$ whereas in the chaotic case they decay exponentially.  We conjecture that this result is universal by using catastrophe theory to show that, on the one hand, the wavefront in the free case has to take the form of an Airy function and its local expansion shows that the power law scaling seen in the numerics holds rigorously, and on the other hand an exponential scaling of the OTOC wavefront must be a signature of nonintegrable dynamics. We find that the crossover between the two regimes is smooth and characterized by an S-shaped curve giving the lifting of Airy nodes as a function of a chaos parameter. This shows that the Airy form is qualitatively stable against weak chaos and consistent with the concept of a quantum Kolmogorov-Arnold-Moser theory. 
\end{abstract}
\maketitle

\textit{Introduction:}	
The hallmark of chaos in classical dynamics is an exponential sensitivity to
small changes in initial conditions (butterfly effect). This is at odds with
quantum mechanics where unitary time evolution means that the overlap between
two states is constant in time. Although quantum systems do not  display chaos,
there are qualitative differences in behavior depending upon whether their
classical limit is integrable or nonintegrable (chaotic) \cite{berry87}. In the
latter case we have `quantum chaos' which is well studied in single-particle
quantum mechanics, including in experiments \cite{jalabert90,marcus92,Milner01,Friedman01,stockmann90,Sridhar91,
Moore94,Steck01,Hensinger01,Chaudhury09,Weinstein02}. 
On the theoretical side, the main approach has
traditionally been through spectral statistics \cite{gutzwiller71,Berry76}.
These have universal properties that depend only on the symmetries of the
Hamiltonian and show close agreement with the predictions of random matrix
theory (RMT)\cite{wigner51,PorterBook,Berry77,Bohigas1984}.
More recently, attention has shifted to many body quantum chaos and
particularly its role in foundational issues such as thermalization in closed
quantum systems. One limitation of RMT is that it does not describe
thermodynamic quantities like temperature and energy that are needed for such
analyzes  \cite{Rigol2015}. This is remedied by the eigenstate thermalization
hypothesis (ETH)
\cite{Deutsch1991,Srednicki1994,Srednicki_1999,Rigol2008,DePalma2015} which has
been numerically verified in a range of generic models
\cite{Leblond2019,Kim2014,Mondaini2016,Kaneko2019} but is violated in
integrable and localized systems
\cite{Biroli2010,Lai2015,RiddellGETH,Ribeiro2013,Vidmar_2016,Nandkishore2015,Iyer2013,Schreiber2015,Luschen2017a,Luschen2017b},
as expected. The ETH generalizes RMT and gives identical predictions if one
focuses on a small enough region of the spectrum.  
Any diagnostic of quantum chaos should therefore clearly differentiate between the integrable
and ETH cases. While the ETH does give rise to the notion of chaotic
eigenstates, it is a time independent statement and does not resemble classical
chaos. In fact, aside from the weak ETH (eigenstate typicality)
\cite{mori2016weak,Brandao2019,Riddell3}, it has no classical counterpart. 

A truly dynamical diagnostic for quantum many body chaos is provided by
out-of-time-ordered correlators (OTOCs) ~\cite{yoshida2019,Swingle2017,Alonso2019,Yan2019,Tuziemski2019,Mao2019,Lewis-Swan2019,Nakamura2019,Belyansky2020,Maldacena2016a}.
They take the form
\begin{equation}
	\label{eq:defotoc}
	C(x,t) = \langle [\hat{A}(t),\hat{B} ]^\dagger  [\hat{A}(t), \hat{B} ] \rangle,
\end{equation}
where $\hat{A}$ and $\hat{B}$ are operators that at $t=0$ only have local
support (act on different individual lattice sites a distance $x$ apart) and hence commute. The
average is usually taken over an ensemble diagonal in the energy basis, but
some studies have considered pure states as well
\cite{Lee2018,Riddell2,Chen2017}.  As  $\hat{A}$ evolves in time, it picks up
weight throughout the lattice, becoming non-local and causing $C(x,t)$ to
become non-zero. This, in effect, tracks the tendency of dynamics to smear
information across the system, and it becomes impossible to determine the
initial conditions from local data alone.  In this respect the OTOC resembles
classical chaos where incomplete information leads to exponential inaccuracy.
Indeed, the late time value of the OTOC in local spin models does appear to be
an indicator of chaos~\cite{Huang2017,Fan2017,Chen2016,Swingle2017,Helu2017,Riddell2,LinOTOCising,Bao2019,Chen2017,Lee2018,Roberts2017,Huang2017v2,McGinley2018}.
In the classical limit commutators become Poisson brackets which are a
diagnostic for classical chaos, and the general expectation is therefore that
OTOCs in nonintegrable models experience exponential growth~ \cite{Maldacena2016a}, 
\begin{equation} \label{eq:simpleexp}
	C(0,t) \sim e^{\lambda_L t} \ ,
\end{equation}
(although we note that integrable systems near unstable points behave similarly~\cite{pappalardi18,hummel19,hashimoto20,cameo20,xu2020,kirkby2021}).
The growth is controlled by the quantum Lyapunov exponent $\lambda_L$ which obeys~\cite{Maldacena2016a}, 
\begin{equation}
	\lambda_L \leq 2 \pi k_B T /\hbar \ .
\end{equation} 
Models that approach the bound are known as fast scramblers.

An OTOC should also display spatial dependence as information propagates across the system.  
A recent conjecture gives the \textit{initial} growth of the OTOC wavefront as  \cite{Khemani20182,Xu2019a,Xu2019b,Xu2019c}
\begin{equation}
\label{eq:SwingleOTOC}
	C(x,t)\sim \exp\left[-\lambda_L\frac{(x/v_B-t)^{1+p}}{t^p}\right]\;.
\end{equation}
This has been tested in several cases and  used to study the many body localization transition~\cite{Nahum2018,Keyserlingk2018,Jian2018,Gu2017,Khemani20182,Xu2019a,Xu2019b,Xu2019c,Sahu2018,Rakovsky2018,Shenker2014a,Patel2017,Chowdhury2017,Jian2018}. When the parameter $p$, or broadening coefficient, takes the value $p=0$, Eq.\ (\ref{eq:SwingleOTOC}) reduces to the simple ``Lyapunov-like" exponential growth of Eq.\ \eqref{eq:simpleexp}, but
for quantum spin models expected to obey ETH it is believed that in general $p>0$ \cite{Khemani20182}. 
However, broadening is not necessarily a general indicator of how close one is to a chaotic model in the sense of ETH \cite{Khemani2018,Anand2020}, 
and puzzles remain concerning the value of $p$ in this early growth regime. 
For example, in two dimensions the values of $p$ coincide in chaotic and integrable  models, so the broadening coefficient
is inadequate for distinguishing them \cite{Khemani20182}, while some studies \cite{Xu2019b,Gopalakrishnan2018,Khemani20182,Le_Doussal_2016,Monthus2006,Kolokolov2008} differ on whether the distinction between values of $p$ even exists in either regime.

For interacting models Eq.\ (\ref{eq:SwingleOTOC}) is usually fitted in regimes where $C (x, t) \lll 1$ \cite{Sahu2018,Xu2019a}, corresponding to times well before the arrival of the main front, where
$C(x,t)$ is exponentially small and not well suited to experimental or numerical verification, as one requires high accuracy. 
We instead focus on the main wavefront region around $x=v_B t$  which is the edge of the OTOC ``light cone''  where $C(x,t) = O(1)$ and show that it carries information about integrability. While there can additionally be signatures of chaos in OTOCs at late times,
including long-time oscillations
\cite{Khemani20182,Fortes2019,Fortes_2020,Anand2020,Wang2020}, it is still preferable
to examine the main front because at late times the signal is more likely to suffer contamination from numerical errors or the environment (in the case of experiments).   

Recent numerical work in free models has shown that the portion of the OTOC around the wavefront is well-fitted by the \textit{leading edge}  of a propagating Gaussian (the peak and trailing edge are not relevant here)
\cite{Riddell2,Riddell4}, 
\begin{equation}  \label{eq:Gaussianwave}
 C_{G}(x,t) \sim e^{-m(x)\left(  x-v_B t \right)^2 +b(x)t} ,
\end{equation}
where  $m(x)$ and $b(x)$ have an inverse power law dependence on $x$.  A Gaussian also occurs in random circuit models \cite{Khemani2018} and wavefront results
suggest it would also be found in the critical Ising model \cite{LinOTOCising}.
In this paper we point out that in free models the wavefront is an example of a \textit{fold catastrophe}, and this allows us to employ arguments from catastrophe theory, a mathematically rigorous theory of bifurcations. A fold arises where two classical solutions (rays) coalesce and is universally dressed by an Airy wavefunction. Local to the wavefront it can be expanded to give precisely the form in  Eq.~\eqref{eq:Gaussianwave} with power law dependence of $m(x)$ and $b(x)$, thereby analytically verifying the numerics of \cite{Riddell2,Riddell4}. However, the converse must also be true: when the scaling disagrees with the catastrophe theory prediction the dynamics must be of a fundamentally different nature to the free case, i.e.\ nonintegrable (chaotic).
We show numerically that the Gaussian wave form of Eq.~\eqref{eq:Gaussianwave} in fact still holds in the chaotic case but that the scaling of $m(x)$ and $b(x)$ is exponential. We conclude that 
in locally interacting models the Gaussian wave form Eq.\ \eqref{eq:Gaussianwave} therefore carries signatures of whether the model is free or ETH-obeying.

\textit{Model:}
We consider a Heisenberg spin Hamiltonian with nearest and next nearest interactions: 
\begin{align}
&\hat{H}(J_1;\Delta;J_2;\gamma)=\hat{H}_{f}+\hat{H}_I\nonumber\\
	&\hat{H}_f =  J_1\sum_{j=1}^{L-1}\left( \hat{S}_j^+ S_{j+1}^- + \text{h.c}\right)\nonumber\\
	&\hat{H}_I =  \Delta \sum_{j=1}^{L-1}\hat{S}_j^Z \hat{S}_{j+1}^Z \nonumber\\
    &+ J_2\sum_{j=1}^{L-2}\left( \hat{S}_j^+ S_{j+2}^- + \text{h.c}\right)+ \gamma  \sum_{j=1}^{L-2}\hat{S}_j^Z \hat{S}_{j+2}^Z\ , 
    \label{eq:Ham}
	\end{align}
and open boundary conditions. 
 This model has free, interacting integrable, and interacting non-integrable regimes depending on the choice of the coefficient vector $ \vec{c}\;\equiv\; (J_1,\Delta,J_2,\gamma)$. We use dimensionless units with $\hbar=k_B=1$ so that time evolution is generated by the unitary operator $U(t) = e^{- i \hat{H} t}$ and thermal states by the density operator $\rho_\beta = \frac{1}{Z} e^{-\beta \hat{H}}$ where $\hat{H} = \hat{H} (J_1;\Delta;J_2;\gamma)$.   
 
 In this work, we focus on two key regimes of our model. The first is the XX chain, $\vec{c}_f = (-0.5,0,0,0)$ which is free.  We choose $J_{1}=-0.5$ because it sets the butterfly velocity to $v_{B}=1$.  The second is $\vec{c}_{\mathrm{ETH}}= (-0.5,1,-0.2,0.5)$ which has been verified to obey the ETH with periodic boundary conditions \cite{Leblond2019}. In this latter case we find $v_{B}>1$, see  the Supplementary Material (SM) \cite{SM}. To explore the transition between these limiting regimes we  also consider intermediate points for which the interactions are turned on via a tunable parameter $0\leq\lambda\leq 1$,
\begin{equation}\label{eq:HamLambda}
\hat{H}_\lambda=\hat{H}_{f}+\lambda \hat{H}_I\; ,
\end{equation}
such that the variable coefficient vector becomes $\vec{c}=(-0.5,\lambda,-0.2 \lambda ,0.5 \lambda)$ and smoothly interpolates between $\vec{c}_f$ and $\vec{c}_{\mathrm{ETH}}$.

When considering the free point $\vec{c}_f$, we use a system size of $L = 1600$, while for the ETH case we use $L = 14$ for the numerics. We leave the investigation of the interacting integrable case to future work. 
In the free case, the numerics are carried out by exactly diagonalizing the model with a Jordan-Wigner transformation, and dynamically evolving the system via the resulting free fermion Hamiltonian \cite{Coleman}. In the ETH case, the numerics are performed with full spectrum exact diagonalization.
We demonstrate that an alternative choice of parameters for $\vec{c}_{\mathrm{ETH}}$ leads to the same basic results in the SM \cite{SM}.

Suitable operators for $\hat{A}(t)$ and $\hat{B}$ must be chosen for the OTOC in Eq.\ (\ref{eq:defotoc}). 
In the ETH regime we use spin operators   $\hat{A}(t) = \sigma_1^Z$, and $\hat{B} = \sigma_m^Z$, where $x$ is the distance between sites 1 and $m$, and 
the average $\langle \ldots \rangle$ is taken over the thermal ensemble restricted to eigenstates with zero magnetization, $m_z $=$ \sum_{j=1}^L  \langle \hat{S}_j^Z \rangle $= 0  and inverse temperature $\beta=1$.
In the free case we perform a Jordan-Wigner transformation from spins to fermions, and for simplicity the OTOC we use in this case is
\begin{equation} \label{eq:FreeOTOC}
		C(x,t) = |a_{m,n}(t)|^2 
\end{equation}
where $a_{m,n}(t) = \{ \hat{f}_m^\dagger(t) , \hat{f}_n \}$.  Here, $\hat{f}_m$ is the annihilation operator for a fermion on site $m$.	
Note that if instead of Eq.\ (\ref{eq:FreeOTOC}) we use Eq.\ (\ref{eq:defotoc}) with operators $\sigma_{m}^z$, 
then in the case of a pure Gaussian state or a thermal ensemble the dominant dynamical term is in fact $|a_{m,n}(t)|^2$, see Refs. \cite{Riddell2,Riddell4} for further details.

\textit{Airy light cones in free systems:} 
In 1972 Lieb and Robinson \cite{lieb72} showed that quantum correlations in spin systems propagate at finite speeds and 
spread out in a light cone-like fashion. 
Pioneering experiments with ultracold atoms and trapped ions \cite{Cheneau2012,Fukuhara2013,langen2013,Jurcevic2014,preiss2015,takasu20}, where a sudden quench leads
to a nonequilibrium state 
\cite{Bakr2009,Weitenberg2011}, have confirmed this behavior.
In particular, the wavefront for interacting bosonic atoms in an optical lattice 
was measured in experiments to have an Airy function profile~\cite{Cheneau2012} in qualitative agreement with theoretical calculations which 
can be done analytically in certain limits~\cite{Barmettler2012}.
The associated problem of domain wall propagation \cite{Hunyadi2004,Eisler2013,Eisler2018,Eisler2020,Kormos2017,Bulchandani2019,Viti2016,Perfetto2017} also
yields Airy functions or related kernels for the wavefront. 
The Airy function shape implies a dynamical scaling behavior, such as a $t^{1/3}$ broadening of the magnetization domain wall in an XX chain \cite{Hunyadi2004}. 
This body of results has led to the notion of an Airy universality class for free systems~\cite{Stephan2019,Fagotti2017,Kiran2020}.

\begin{figure}[t!]
		\centering
		\includegraphics[width=\linewidth]{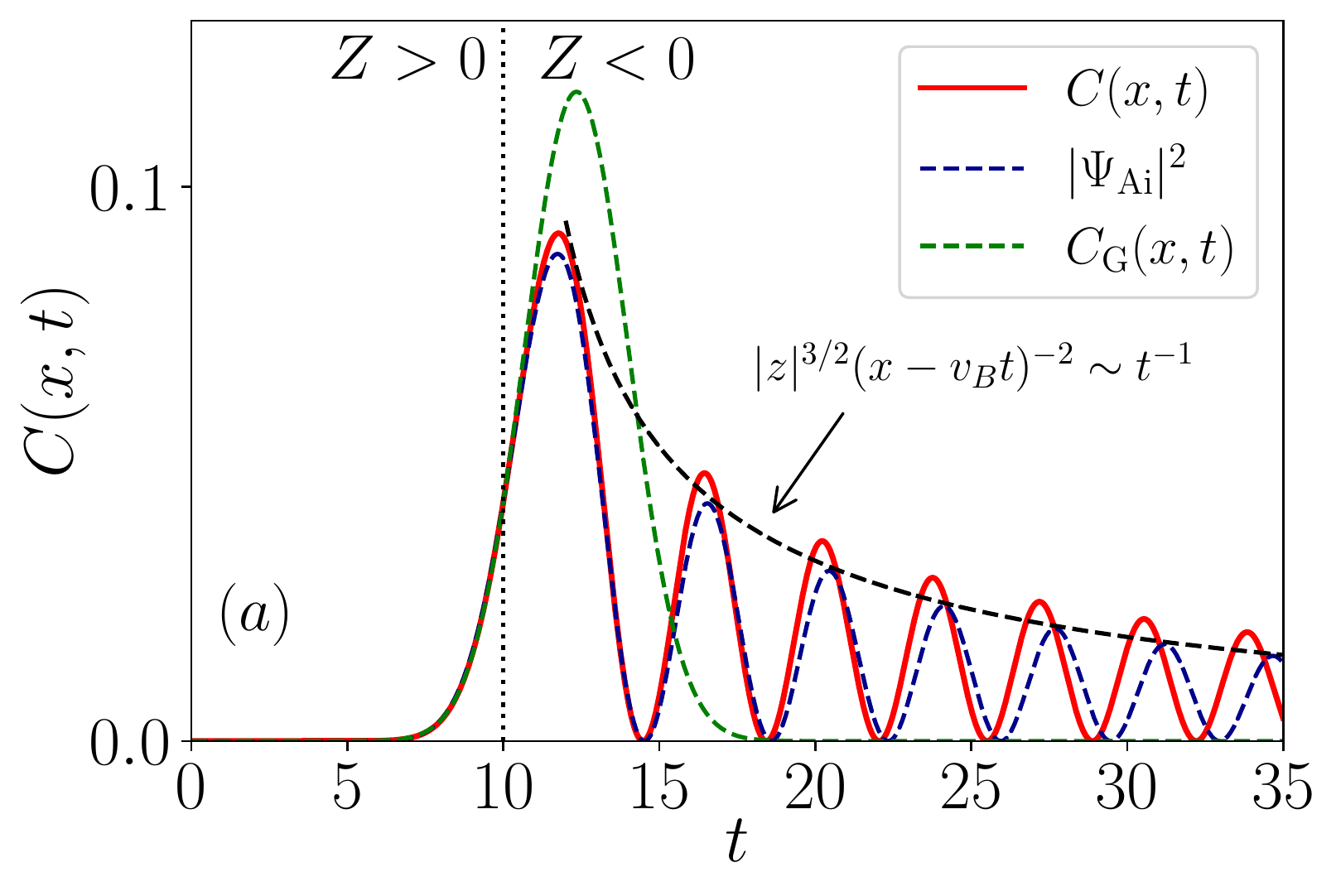}
		\includegraphics[width=\linewidth]{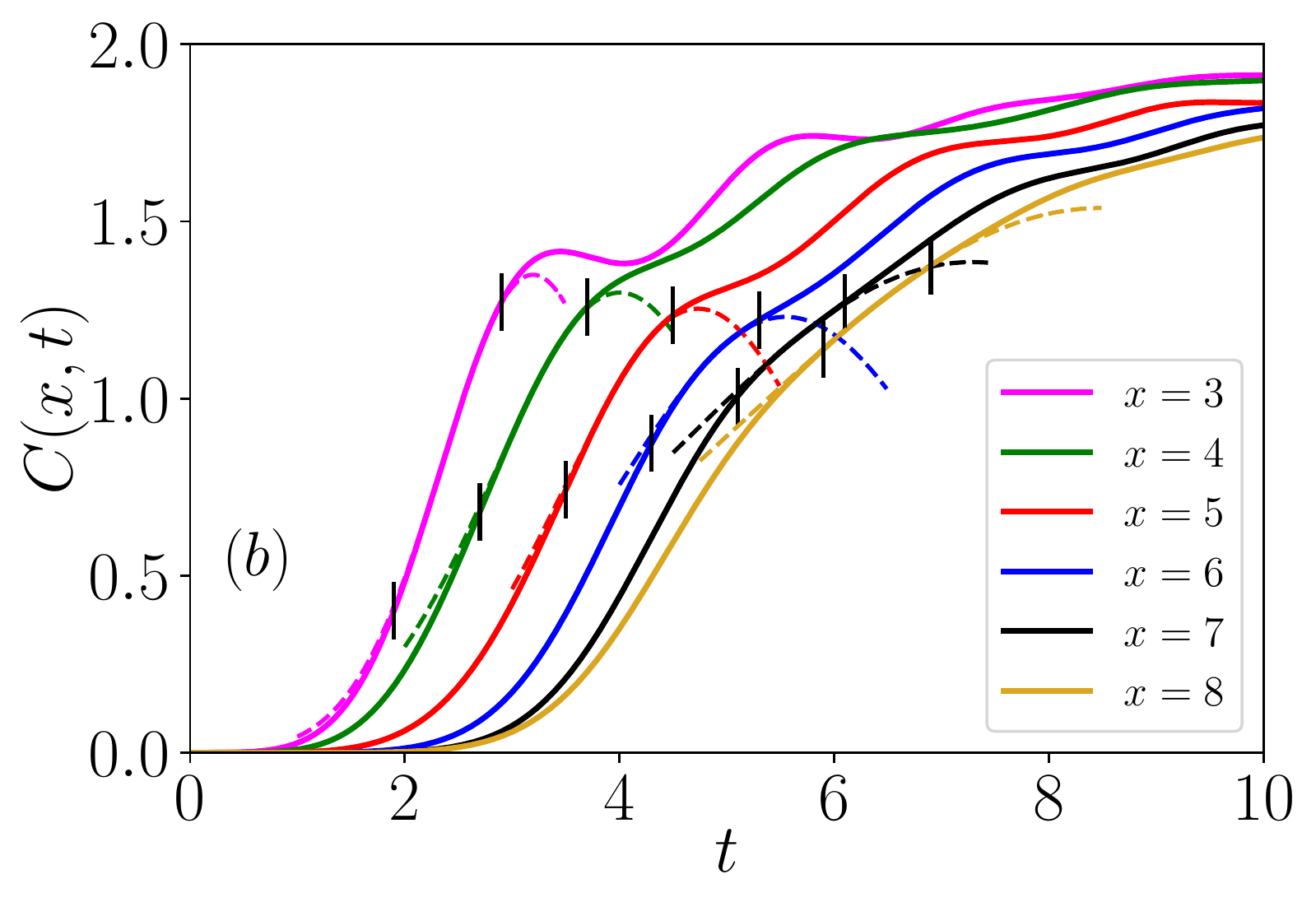}
		\caption{Wavefronts of $C(x,t)$
                \textbf{(a)} Free case, $\vec{c}_{f}$. Here $x=10$. Exact numerical results (solid red) along with a fit to the Gaussian form Eq.~(\ref{eq:Gaussianwave}) (green), and the Airy result from Eq.\ (\ref{eq:airy}) (blue). The expected $t^{-1}$ decay in the amplitude is also shown (black).   Note that the part of the wavefront we are fitting to a Gaussian is only a window around $t=10$.
                \textbf{(b)} ETH case, $\vec{c}_{\mathrm{ETH}}$, at different positions $x$. Solid lines indicate the exact OTOC data, and dashed lines are fits to Eq.\ \eqref{eq:Gaussianwave} centered at $x=v_Bt$.  Vertical bars indicate the fitting window given in Eq.\ \eqref{eq:fitwindow}.
            \label{fig:airy_gaussian}}
	\end{figure}	

To explain the ubiquitous presence of Airy functions we use catastrophe theory. Due to topological properties, \textit{a finite set} of bifurcations are structurally stable meaning they are robust to deformations and perturbations and hence occur generically in nature without the need for fine tuning; they are classified by catastrophe theory and form a hierarchy where the higher catastrophes contain the lower ones~\cite{thom75,arnold75,Poston&Stewart,saunders80}. In optics and hydrodynamics catastrophes appear as \textit{caustics} where the amplitude diverges in the classical limit, with examples including rainbows, gravitational lensing, ship's wakes and rogue waves \cite{airy1838,Nyebook,kelvin1905,hohmann2010,Onoratoab2013}.
Each class of catastrophe is specified by a normal form $\Phi(\mathbf{s},\mathbf{C})$
which is a polynomial in the state variables $\mathbf{s}=\{s_{1},s_{2},\ldots \}$ that specify the rays and linear in the control parameters $\mathbf{C}=\{C_{1},C_{2},\ldots\}$ that include the coordinates and other parameters. In physical terms $\Phi(\mathbf{s},\mathbf{C})$ is the action and the local wavefunction  is obtained by an elementary path integral over all configurations \cite{duistermaat1974,guillemin1977,Kirkby2022}
\begin{equation}
\label{eq:wavecatastrophe}
 \Psi(\mathbf{C})=\left(\frac{1}{2 \pi}\right)^{n/2} \int_{-\infty}^{\infty} \cdots \int d^{n}s  \ e^{i \Phi(\mathbf{s};\mathbf{C})} \ .
\end{equation}  
The simplest in the hierarchy is the fold catastrophe which has one state variable $s$, one control parameter $Z$,  and a cubic action $\Phi=s^3/3+Zs$. In this case the wavefunction is directly proportional to the Airy function $\Psi(Z)=\sqrt{2 \pi} \mathrm{Ai}(Z)$ which is plotted in  Fig.\  \ref{fig:airy_gaussian}(a).
The two extrema of $\Phi(s,Z)$ describe the coalescence  of two rays as a function of $Z$. The next catastrophe in the hierarchy is the cusp which has a quartic action (coalescence of three rays) and the wavefunction is known as the Pearcey function \cite{Pearcey1946}.

The connection to OTOCs is that in the integrable case where  spins map to noninteracting fermions we can show that the edge of a light cone is a catastrophe/caustic where the number of saddles of the action changes, corresponding to coalescing quasiparticle trajectories  \cite{Kirkby2019}. Consider the case where a quench excites a Bogoliubov fermion at the site at $x=0$. The resulting wavefunction is \cite{SM}
\begin{align}
\Psi(x,t)= & \braket{x|\mathrm{e}^{-\mathrm{i}\hat{H}t} \, \hat{b}_{x=0}^\dagger|0}=\;\bra{x}\sum_{k}\mathrm{e}^{-\mathrm{i}\epsilon(k)t}\ket{k} \nonumber \\ & \approx\frac{\sqrt{a}}{2\pi}\int_{-\frac{\pi}{a}}^{\frac{\pi}{a}}\mathrm{d}k\;\mathrm{e}^{\mathrm{i} [kx-\epsilon(k)t] } \; \label{eq:wavefunc1}
\end{align}
where $a$ is the lattice constant.  The operators $\hat{b}_{x}$ are the linear combinations of $\hat{f}_{m}$ and $\hat{f}_{m}^{\dag}$ that diagonalize the Hamiltonian via a Bogoliubov transformation and $\epsilon(k)$ is the Bogoliubov dispersion relation [for the XX chain $\epsilon(k)=2 J_{1} \cos k a$]. Putting  $\Phi(k,x,t) = kx-\epsilon(k)t$, a caustic occurs at  quasimomentum $k_{c}$ where two conditions are satisfied \cite{berry81}
\begin{equation}(\partial \Phi/\partial k)_{k_c}=0 \quad \mbox{and} \quad (\partial^2\Phi/\partial k^2)_{k_c}=0. \label{eq:caustic_conditions}
\end{equation}
The first is Fermat's principle that gives classical rays as saddles of the action $\Phi=\int L dt$ where $L=k\dot{x}-\epsilon(k)$, and the second defines the caustic as the locus of points $(x,t)$ where saddles coalesce. 
Together, these conditions correspond exactly to the Lieb-Robinson (LR) bound for a light
cone as being determined by the maximum value of the group velocity $d
\epsilon/d k$ of the fermions \cite{Kirkby2019,stephan11,calabrese2012},
$v_{\mathrm{LR}}= \max_{k} |d \epsilon/d k|$. 

Why do caustics occur? In classical integrable systems trajectories are confined to live on invariant tori of dimension $N$ in phase space of dimension $2N$, where $N$ is the number of degrees of freedom. Each torus is associated with a family of trajectories that wrap around the torus either periodically or quasiperiodically. When projected down onto coordinate space the edges of the torus lead to a diverging density of trajectories that all have the same position (but different momenta), i.e.\ a caustic in coordinate space \cite{berry83Houches}.

Identifying light cones in integrable systems as caustics allows rigorous results from catastrophe theory to be applied: 
\textbf{i)} The only structurally stable bifurcations in two dimensions (the space-time formed by $x$ and $t$) are fold lines that meet at cusp points, as someone who has ironed a shirt knows;
\textbf{ii)} For a
fold catastrophe the phase $\Phi(s,Z)$ is cubic in $s$; 
\textbf{iii)} 
There exists a diffeomorphism from the physical variables $(k,x,t)$ to the canonical Airy cubic form $(s,Z)$. Therefore, a Taylor expansion truncated at third order about the caustic can give the exact wavefunction in the neighborhood of that point. Adding higher order terms will not affect the qualitative behavior because the merging of two stationary points is fully captured by a cubic action with the tunable parameter $Z$. 

Performing the transformation of variables $s^3=2(k-k_c)^3/[t\partial^3_k\epsilon(k_c) ]$ in Eq.\ (\ref{eq:wavefunc1})
gives \cite{Xu2019b,Khemani20182}
\begin{equation}
		\Psi_{\mathrm{Ai}}(x,t) \sim \sqrt{a} \left(\frac{-2}{\partial^3_k\epsilon(k_c)t}\right)^{1/3}\mathrm{e}^{\mathrm{i} \Phi(k_c, x,t)} \; \mathrm{Ai}(Z) \;,
		\label{eq:airy}
\end{equation}
	where $Z = (x-v_B t)\left\vert t\partial^3_k\epsilon(k_c)/2\right\vert^{-1/3}$ \cite{SM}. In Fig.~\ref{fig:airy_gaussian}(a), we plot $|\Psi_{\mathrm{Ai}}(x,t)|^2$
alongside the numerical result at the point $x=10$, with the caustic at $Z=0$ marked by the vertical dotted line. The Airy wavefunction gradually goes out
of phase at longer times because the Taylor expansion was made at a single
point, but the range could be extended via a uniform mapping onto the tail of a WKB solution \cite{berry72}. From
the asymptotics of the Airy function as $Z \rightarrow - \infty$ it follows
that the amplitude of the OTOC decays as $\vert Z \vert^{3/2}/(x-vt)^2  \sim 1/t$ inside the light cone
(in agreement with Refs.\ \cite{LinOTOCising,Bao2019}), and the fringe spacing becomes {\it constant}.
Travelling along the wavefront $x/t=v_B$ one finds that the amplitude decays as $x^{-2/3}$ and the width of the primary fringe grows as $t^{1/3}$.
Furthermore, 
Eq.\ (\ref{eq:airy}) also correctly predicts the early time growth: keeping
just the first term of the $Z\to\infty$ asymptotic series for the Airy function
\cite{SM} gives the universal $p=1/2$ form of the OTOC in Eq.\
\eqref{eq:SwingleOTOC} \cite{Xu2019b,Xu2019a,Khemani20182}.

Airy functions have been derived for OTOCs before \cite{Xu2019b,Khemani20182,Kiran2020}. Our point here is that catastrophe theory guarantees that these results are exact, providing the assumptions underlying Eqns.\ (\ref{eq:wavefunc1}) and (\ref{eq:caustic_conditions}) hold, namely noninteracting quasiparticles with dispersion $\epsilon(k)$. But if these results are exact, any deviation would imply that the assumption of free quasiparticles must be breaking down. We turn to this case below.  

The fold is only the first in a hierarchy, and in fact, the two edges of the light cone should generically meet at a cusp. However, the high symmetry of the XX model means that $\epsilon(k)$ is so simple that only two rays can coalesce at once and no cusp occurs, just two pure fold lines that meet at $x=t=0$. If a symmetry breaking term is added (like in the XY model) three rays can coalecse at the origin and the
back-to-back Airy functions are locally replaced by a Pearcey function~\cite{Kirkby2019}.

\begin{figure}[t!]
	\centering
	\includegraphics[width=\linewidth]{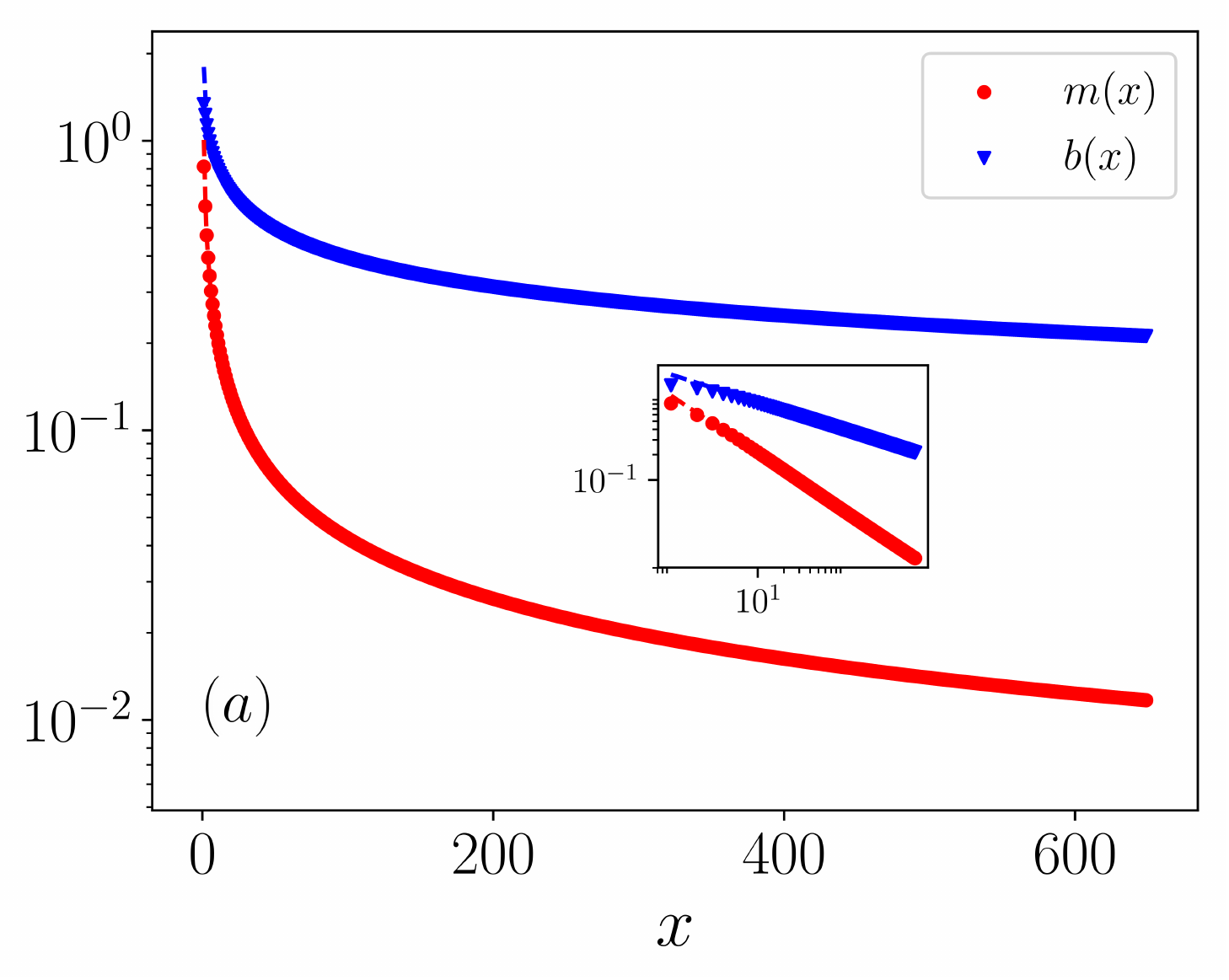}
	\includegraphics[width=\linewidth]{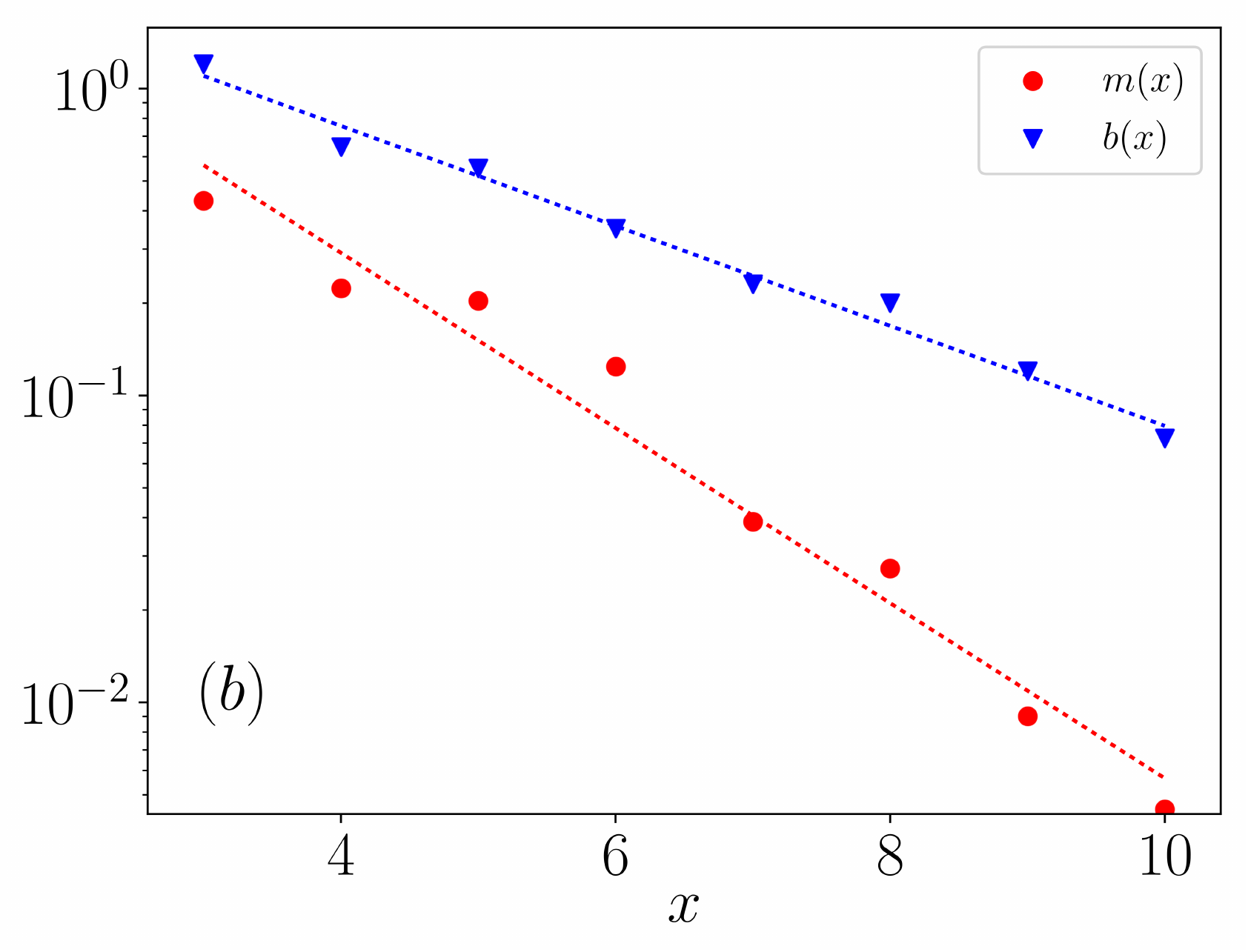}
        \caption{Log-linear plots for the Gaussian parameters $m(x)$ [red] and $b(x)$ [blue]. 
        \textbf{(a)} The free Hamiltonian, $\vec{c}_f$. Dashed lines are fits to Eq.\ \ref{eq:freescale}. Inset is a log-log plot of the same data. 
        \textbf{(b)} The ETH Hamiltonian, $\vec{c}_{\mathrm{ETH}}$, using the same data as in Fig.\ \ref{fig:airy_gaussian}(b) indicating exponential decay of $m(x)$, $b(x)$.} 
	\label{fig:scaling}
\end{figure}

\textit{Profile of the wavefront in the ETH case:} 
In Fig.~\ref{fig:airy_gaussian}(b) we plot the exact results for the OTOC for $\vec{c}_{\mathrm{ETH}}$.  
Fringes are partially visible at smaller $x$ but the Airy nodes have disappeared. At $x=3$ the wavefront has quite a sharp slope, indicating that the process of scrambling (the increase in non-locality of the observable) is still in full swing. By $x=8$, 
the slope of the OTOC at the wavefront has significantly decreased.
The Gaussian waveform of Eq.\ \eqref{eq:Gaussianwave} provides an excellent local fit to the wavefront in both the free~\cite{Riddell2,Riddell4} and 
chaotic regimes, as seen from the dashed curves in Fig.~\ref{fig:airy_gaussian}(a) and \ref{fig:airy_gaussian}(b), respectively. 
The fit is performed over the range 
\begin{equation}
    t=\frac{x}{v_B} \pm \Delta t,\label{eq:fitwindow}
\end{equation}
where $\Delta t\approx 0.5$ 
gives a reasonably large window to describe the shape of $C(x,t)$ at the wavefront.  
The fits for the parameters $m(x)$ and $b(x)$ in Eq.\ \eqref{eq:Gaussianwave} are shown in Fig.\ \ref{fig:scaling} and indicate strong agreement with the data: 
errors on each term are on the order of $10^{-7}$ to $10^{-9}$ for all $x$. 
A crucial ingredient to identify the parameters in the ETH case is to first determine the butterfly velocity $v_B$, 
which can be done using velocity-dependent Lyapunov exponents \cite{Khemani20182,zhang2020}, as demonstrated in the SM \cite{SM}. 
We find that the velocity for the ETH model characterized by $\vec{c}_{\mathrm{ETH}}$ is roughly $v_B \approx 1.28$ (in contrast to $v_B = 1$ for $\vec{c}_{f}$). 
Although the free and ETH wavefronts both display flattening, the scaling properties of $m(x)$ and $b(x)$ are fundamentally different in the two regimes as we now show.

\textit{Scaling in Free Models:}
By expanding the Airy wavefunction given in Eq.\ (\ref{eq:airy}) about the caustic at $z=0$ we obtain 
\begin{equation}
 m(x)=\;\frac{c_m}{x^{\frac{2}{3}}} \; , \enspace b(x)=\;\frac{c_b}{x^{\frac{1}{3}}}\label{eq:FreeOTOCGaussian_b}	\;,
\end{equation}
where $c_m$ and $c_b$ are constants that depend explicitly on the dispersion relation (see the SM \cite{SM} for details). 
Due to the universality of the Airy wavefunction, this scaling is expected to hold for models which can be written in terms of freely propagating quasiparticles.	 Furthermore, corrections beyond quadratic order in $x-v_Bt$ can be obtained. However, the cubic term in the exponent falls off rapidly (at least as $x^{-1}$), and so it is reasonable, even at moderate distances, to keep only the Gaussian approximation. 
We have numerically verified Eq.~\eqref{eq:FreeOTOCGaussian_b} and the results are shown in Fig.~\ref{fig:scaling}(a). Fitting the scaling of each parameter for distances $0<x\leq650$ we find,
\begin{equation} \label{eq:freescale}
m(x) \propto \frac{1}{x^{a_m}}\;, \enspace b(x) \propto \frac{1}{x^{a_b}} \;,
\end{equation}
with $a_m = 0.68857 \pm 0.00008$, and $a_b = 0.33043 \pm 0.00002$,  indicating good agreement with the expected values.
We also note that because $m(x) \propto b(x)^2$, $m(x)$ falls off significantly quicker than $b(x)$. 
This may point to an intermediate regime in $x$ where the OTOC is well described by $C(x,t) \sim e^{b(x) t}$.

\begin{figure}[t]
	\centering
    \includegraphics[width=\linewidth]{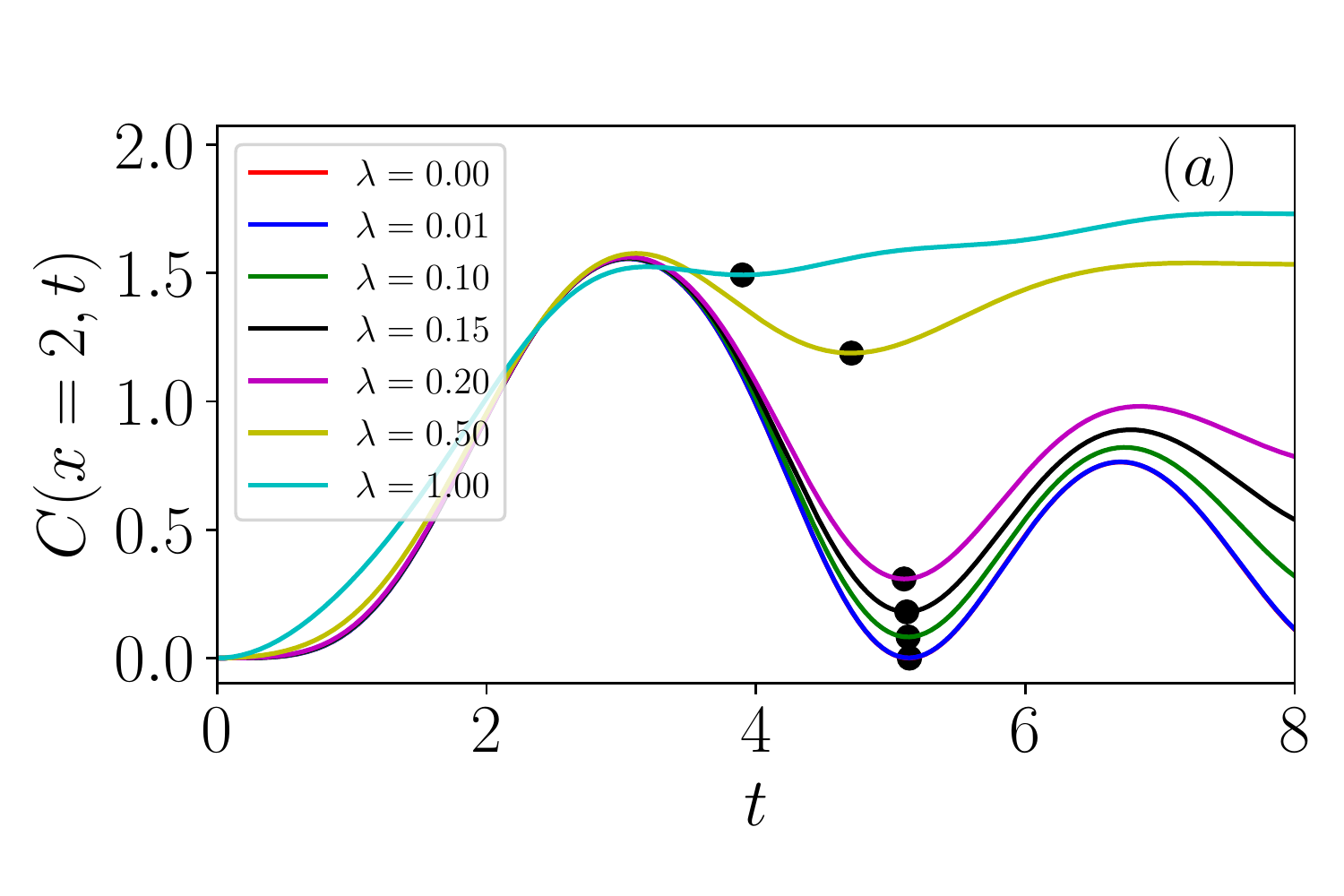}
	\includegraphics[width=0.97\linewidth]{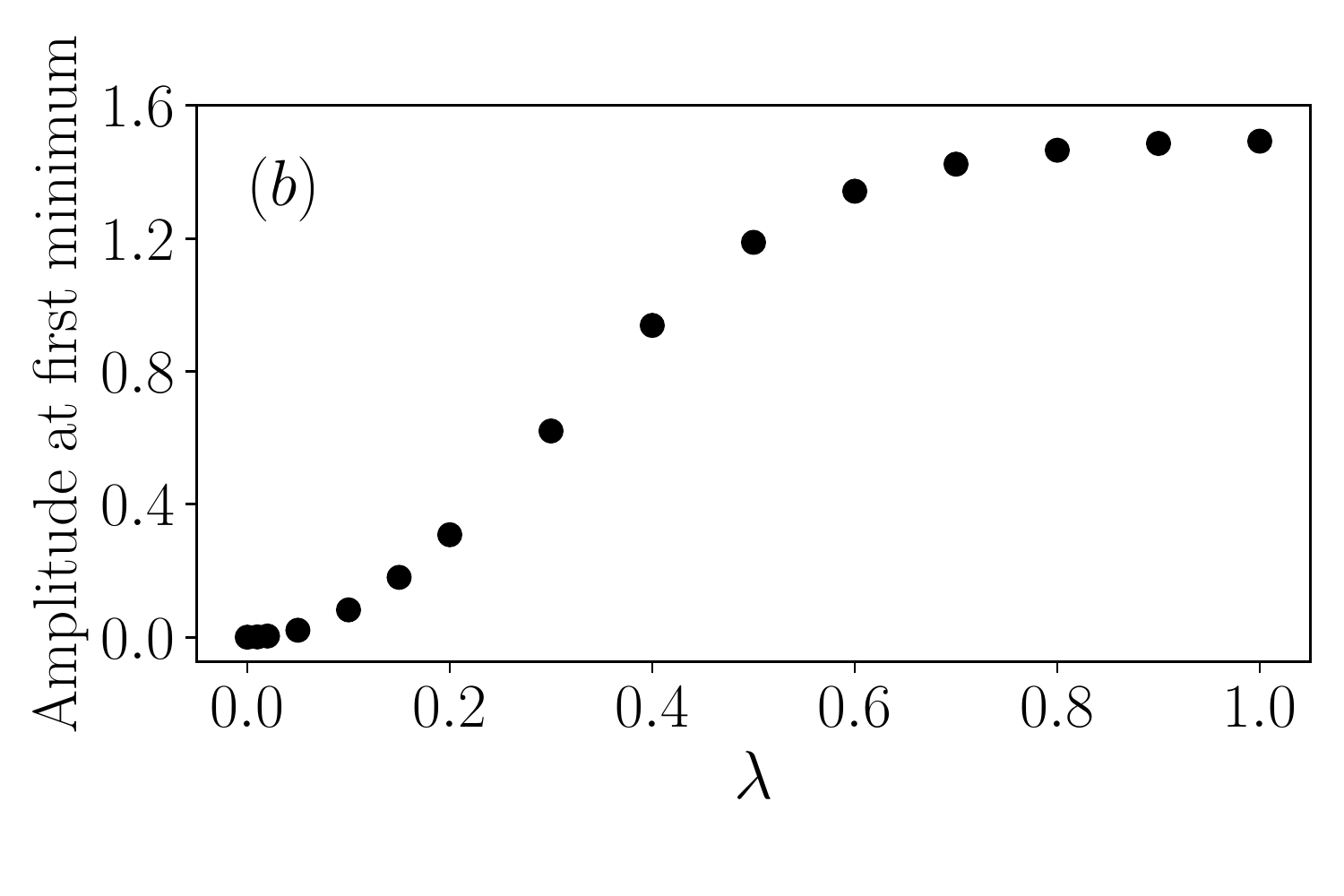}
	\caption{\textbf{(a)} Tracking of the first OTOC minimum as a function of time.  Interactions are introduced by varying  $\vec{c}(\lambda) = (-0.5,\lambda, -0.2 \lambda, 0.5 \lambda )$,  such that $\Vec{c}(0) = \vec{c}_f$ and $\Vec{c}(1) = \Vec{c}_{\mathrm{ETH}}$. Black dots indicate the first minimum after the wave front. \textbf{(b)}  The height of the first minimum as a function of $\lambda$ follows an S-shaped curve. The transition to chaos appears to be smooth and the initial lift off is slow indicating that the Airy function solution is qualitatively stable against weak chaos.} 
	\label{fig:velocity2}
\end{figure}

\textit{Scaling in ETH regime:}
In Fig.\ \ref{fig:scaling}(b) we show a plot of the data for the
$\vec{c}_{\mathrm{ETH}}$  case. 
A linear trend emerges,
implying that the spatial dependence on $m(x)$ and $b(x)$ in the ETH regime
exhibits exponential rather than power-law decay, 
\begin{equation}
		b(x) \sim e^{-c x}, \enspace m(x) \sim e^{-w x},
\end{equation}
where $c,w>0$ are constants. We find that $c = 0.38 \pm 0.02$ and $w = 0.66 \pm
0.05$. 
Like the free case, 
$m(x) \propto b(x)^2$, 
however, as shown in the SM \cite{SM}, this is not generally the case.
	
The exponentially decaying behavior of $m(x)$ and $b(x)$ is clearly distinct
from the free fermion case. 
This indicates that the Gaussian waveform can
distinguish ETH-obeying
from free dynamics. 
In both Figs. \ref{fig:scaling}(a) and (b)
$m(x), b(x)$ decay by upwards of two orders of magnitude as a function of
position, however the exponential decay in the ETH regime ensures that this
occurs over a short distance of $x \approx 10$ while in the free model it takes
a distance of $x \approx 600$. Thus, the general flattening of the OTOC at the
wavefront (see e.g. Fig.\ \ref{fig:airy_gaussian}) occurs {\it much} faster in
thermalizing models. 
	
 \textit{Crossover between free and ETH regimes:}
Fig.~\ref{fig:velocity2}(a) shows the OTOC as a function of time at the fixed coordinate $x=2$ for the system described by the tunable Hamiltonian in Eq.\ \eqref{eq:HamLambda} for a range of $\lambda$. For the free system ($\lambda=0$), the local minima correspond to Airy zeroes. As $\lambda$ is increased the integrability is broken and the nodes are lifted up from zero although they remain local minima even in the ETH regime. The amplitude at the first minimum as a function of $\lambda$ is plotted in Fig.\ \ref{fig:velocity2}(b) and follows a characteristic S-shaped curve that saturates in the ETH regime. The monotonic increase of this amplitude as a function of $\lambda$ means that it provides a simple measure of the effect of chaos on the OTOC, and from this point of view it is notable that the lift off is initially slow. A gradual lifting rather than immediate destruction of nodes indicates that the Airy description of the light cone remains approximately correct even in the presence of small integrability-breaking terms. Given the association between caustics and invariant tori in phase space, we are reminded of the Kolmogorov–Arnold–Moser (KAM) theorem that ensures some tori (and therefore caustics) persist for weakly chaotic systems \cite{arnoldbook}. The survival of the Airy-like form of the wavefront with $\lambda>0$ is suggestive of the existence of a quantum KAM theory \cite{Hose83,brandino15,Bastianello21,Burgarth21,Bulchandani2022}.

\textit{Conclusions:}  
 Both numerics and rigorous results based on catastrophe theory show that close to the wavefront free and ETH models can be distinguished by the difference in scaling of the parameters $m(x)$, $b(x)$ in Eq.~(\ref{eq:Gaussianwave}). 
The ability of modern experiments to  realize spin models and measure light cone profiles \cite{Cheneau2012,Fukuhara2013,langen2013,Jurcevic2014,preiss2015,takasu20} holds out the possibility that this prediction can be tested in the laboratory. Our results also demonstrate the structural stability of the Airy function wavefront against weak chaos. The connection between Airy functions, caustics, and invariant tori in phase space, suggests there is a quantum analogue of the KAM theory \cite{brandino15}. Such a theory would imply that the transition to chaos is smooth and is consistent with the widespread prediction and observation of prethermalized (i.e.\ nonthermal) phases at short to intermediate timescales after quenches and before full thermalization sets in \cite{Kinoshita2006,gring12,Gaunt2013,Langen2016,Bertini2016,Rauer2018,Choi2019,Rubio2020,Ueda2020,Kurlov2022}.

\begin{acknowledgments}
We acknowledge the support of
the Natural Sciences and Engineering Research Council of Canada (NSERC) through Discovery
Grants (No. RGPIN-2017-05759 and No. RGPIN-2017-
06605).
This research was enabled in part by support provided by SHARCNET (sharcnet.ca) and the Digital Research Alliance of Canada (alliancecan.ca).
\end{acknowledgments}

        \twocolumngrid



%

\end{document}